\documentclass[galaxies,article,accept,moreauthors,10pt]{mdpi}

\pdfoutput=1

\usepackage{graphicx,amsmath}
\usepackage{booktabs}
\usepackage{multirow}
\usepackage{soul}
\usepackage{microtype}

\allowdisplaybreaks

\newcommand{\approptoinn}[2]{\mathrel{\vcenter{
  \offinterlineskip\halign{\hfil$##$\cr
    #1\propto\cr\noalign{\kern1.5pt}#1\sim\cr\noalign{\kern-1.5pt}}}}}
\newcommand{\appropto}{\mathpalette\approptoinn\relax}

\firstpage{1}
\articlenumber{x}
\doinum{10.3390/------}
\pubvolume{6}
\pubyear{2018}
\copyrightyear{2018}
\history{Received: 30 January 2018; Accepted: 19 March 2018; Published: date}

\Title{Applying MOG to Lensing: Einstein Rings, Abell 520 and the Bullet Cluster}

\Author{J. W. Moffat $^{1,2}$, S. Rahvar $^{1,3}$ and V. T. Toth $^{1,}$*}%
\address{%
$^{1}$ \quad Perimeter Institute for Theoretical Physics, Waterloo, ON N2L 2Y5, Canada; jmoffat@perimeterinstitute.ca~(J.W.M.); sohrabrahvar@gmail.com (S.R.)\\
$^{2}$ \quad Department of Physics and Astronomy, University of Waterloo, Waterloo, ON N2L 3G1, Canada\\
$^{3}$ \quad Department of Physics, Sharif University of Technology, PO Box 11365-9161 Tehran, Iran}
\corres{Correspondence: vttoth@vttoth.com}

\abstract{We investigate gravitational lensing in the context of the MOG modified theory of gravity. Using a formulation of the theory with no adjustable or fitted parameters, we present the MOG equations of motion for slow, nonrelativistic test particles and for ultrarelativistic test particles, such as rays of light. We demonstrate how the MOG prediction for the bending of light can be applied to astronomical observations. Our investigation first focuses on a small set of strong lensing observations where the properties of the lensing objects are found to be consistent with the predictions of the theory. We also present an analysis of the colliding clusters 1E0657-558 (known also as the Bullet Cluster) and Abell~520; in both cases, the predictions of the MOG theory are in good agreement with observation.}


\begin{document}

\section{Introduction}

Ever since Einstein's 1915 prediction of the angle by which a ray of light is bent by the eclipsed Sun, and its spectacular (albeit now seen as somewhat controversial) confirmation by observations led by Arthur Eddington during the 1919 solar eclipse \cite{Eddington1920}, the effects of gravity on light have been viewed as a mechanism by which the predictions of a gravity theory like general relativity can be tested directly.

This remains the case today. Indeed, gravitational lensing is viewed as a possible means to distinguish between competing theories of gravitation and cosmology: in particular, between theories of modified gravity versus the collisionless cold dark matter (CDM) that forms the basis of the $\Lambda$CDM ``standard model'' of cosmology.

The discovery of the colliding clusters 1E0657-558, also known colloquially as the Bullet Cluster, was~initially viewed by many as a decisive case in favor of a CDM model \cite{Clowe2004}, as the lensing associated with cluster components depleted in gas (normally, the dominant baryonic component) seemed inconsistent with theories of modified gravity. More recently, however, the colliding clusters Abell~520 \cite{Mahdavi2007,Jee2012} appeared as a counterexample. The lensing associated with gas-rich core regions of this object that contain no visible galaxies can only be understood if the presence of some dark matter is assumed, which is not consistent with the collisionless nature of CDM in the standard model.

The gravitational lensing maps obtained from the observation of colliding clusters are examples of weak lensing. There are also cases of strong lensing observed in the deep sky. These come in the form of Einstein rings, a visually stunning phenomena where the image of a background object is heavily distorted, forming a partial or full ring around a foreground object as a result of the latter's gravity. Measuring the angular size of the ring can be compared to the observed distances (obtained from redshifts) to the foreground and background objects, from which one can make deductions about the size of the lensing~mass.

Our Modified Gravity (MOG) theory, also known as Scalar-Tensor-Vector Gravity (STVG) \cite{Moffat2006a,Moffat2007e}, is a classical theory of gravity based on a Lagrangian principle. Phenomenologically, the theory comprises metric gravity with a variable gravitational constant greater than Newton's, and a repulsive vector field of finite range that partially cancels out the attractive force. The gravitational constant and the mass of the vector field are themselves promoted to scalar fields. This theory is known to yield results that are consistent with precision solar system experiments. It was applied successfully to a sample of over \mbox{100 spiral galaxies}, correctly modeling their detailed rotation curves without dark matter \cite{Brownstein2006a,Brownstein:2009gz}. Furthermore, the theory was also used for cosmological investigations, successfully modeling the acoustic power spectrum of the cosmic microwave background and the galaxy-galaxy power spectrum \cite{Moffat2007e}.

The MOG theory was also used to model the weak gravitational lensing of the Bullet Cluster \cite{Brownstein2007,Israel2016}. The predictions of MOG were found to be consistent with observation, demonstrating that it is, in fact, possible for a modified gravity theory to yield agreement with the Bullet Cluster.

In this paper, we first present a condensed review of the MOG theory, focusing on the development of the weak field, low velocity equations motion on the one hand, and the equations governing ultrarelativistic particles on the other. This material is presented in Section~\ref{sec:theory}. In Section~\ref{sec:rings}, we apply the theory to a small set of strong lensing objects for which detailed observational data are available. In Section~\ref{sec:clusters}, we investigate the Bullet Cluster and Abell~520. This section contains our main result, demonstrating that the behavior of both clusters is consistent with the MOG theory. Finally, in Section~\ref{sec:conclusions} we present our conclusions and~outlook.

In this paper, we use the following conventions: the symbol $\nabla_\mu$ is used to denote covariant differentiation with respect to the metric $g^{\mu\nu}$, while the symbols $R$, $\Lambda$, and $g$ represent the Ricci scalar, the~cosmological constant, and the determinant of the metric tensor, respectively. We define the Ricci tensor as $R_{\mu\nu}=\partial_\alpha\Gamma^\alpha_{\mu\nu}-\partial_\nu\Gamma^\alpha_{\mu\alpha}+\Gamma^\alpha_{\mu\nu}\Gamma^\beta_{\alpha\beta}-\Gamma^\alpha_{\mu\beta}\Gamma^\beta_{\alpha\nu}$.
Unless otherwise noted, we set the speed of light, $c=1$; we use the metric signature $(+,-,-,-)$.

\section{The MOG Theory}
\label{sec:theory}

In this section, we summarize the main results of MOG as they apply to the bending of light. The~presentation here is a condensed version of that given by~\cite{Moffat2008a}.

\subsection{Lagrangian Formulation}
\label{sec:formulation}

Our modified gravity theory is based on postulating the existence of a massive vector field, $\phi_\mu$. The~choice of a massive vector field is motivated by our desire to introduce a {\em repulsive} modification of the law of gravitation at short range. The vector field is coupled universally to matter. The theory, therefore, has three constants: in addition to the gravitational constant $G$, 
a further constant $\mu$ 
arises as a result of considering a vector field of non-zero mass, and controls the coupling range. The theory promotes $G$ and $\mu$ to scalar fields, hence they are allowed to run, resulting in the following action \cite{Moffat2006a,Moffat2007e}:
\begin{equation}
S=S_G+S_\phi+S_S+S_M,
\end{equation}
where
\begin{align}
S_G=&-\frac{1}{16\pi}\int\frac{1}{G}\left(R+2\Lambda\right)\sqrt{-g}~d^4x,\\
S_\phi=&-
\int\left[\frac{1}{4}B^{\mu\nu}B_{\mu\nu}-\frac{1}{2}\mu^2\phi_\mu\phi^\mu+V_\phi(\phi)\right]\sqrt{-g}~d^4x,\\
S_S=&-\int\frac{1}{G}\Big[\frac{1}{2}g^{\mu\nu}\left(\frac{\nabla_\mu G\nabla_\nu G}{G^2}+\frac{\nabla_\mu\mu\nabla_\nu\mu}{\mu^2}\right) 
+\frac{V_G(G)}{G^2}+\frac{V_\mu(\mu)}{\mu^2}\Big]\sqrt{-g}~d^4x,
\end{align}
where $S_M$ is the ``matter'' action, $B_{\mu\nu}=\partial_\mu\phi_\nu-\partial_\nu\phi_\mu$, while $V_\phi(\phi)$, $V_G(G)$ and $V_\mu(\mu)$ denote the self-interaction potentials associated with the vector field and the two scalar fields. 

In the case of a spherically symmetric field in vacuum around a compact (point) source, we were able to derive an exact numerical solution \cite{Moffat2007e}. We found that the scalar fields $G$ and $\mu$ remain constant except in the immediate vicinity of the source. The spatial part of the vector field $\phi_\mu$ is zero, while its $t$-component obeys a simple exponential relationship. Meanwhile, the metric is approximately the Reissner-Nordstr\"om metric of a source with a fifth-force charge.

Specifically, given a spherically symmetric, static metric in the standard form
\begin{equation}
d\tau^2=Bdt^2-Adr^2-r^2(d\theta^2+\sin^2\theta d\phi^2),\label{eq:schwarzschild}
\end{equation}
\textls[-20]{we found \cite{Moffat2006a,Moffat2007e} that, for a source mass $M$, the field equations yield a solution similar to the Reissner-Nordstr\"om metric of a charged black hole, the charge in this case being the fifth-force charge associated with the source~mass:}
\begin{align}
A&\simeq B^{-1},\\
B&\simeq 1-\frac{2G_0M}{r}+\frac{
G_0Q_5^2}{r^2},\label{eq:BB}\\
G&\simeq G_0=G_N+(G_\infty-G_N)\frac{M}{(\sqrt{M}+E)^2},\label{eq:G}\\
\mu&\simeq \mu_0=\frac{D}{\sqrt{M}},\label{eq:mu}\\
\phi_t&\simeq -Q_5\frac{e^{-\mu r}}{r},
\end{align}
where $G_N$ is Newton's constant of gravitation, $Q_5=\kappa M$ is the fifth force charge associated with the source mass $M$, while $G_\infty$, $D$ and $E$ are constants of integration. Further, a comprehensive fit to a variety of astronomical and cosmological data sets (including galaxy rotation curves, the large-scale distribution of matter, and peaks of the microwave background angular power spectrum) yielded the following approximate values \cite{Moffat2007e}:
\begin{align}
\kappa&= \sqrt{G_N},\\
D&\simeq 6250~M_\odot^{1/2}\mathrm{kpc}^{-1},\\
E&\simeq 25,000~M_\odot^{1/2},\\
G_\infty&\simeq 20G_N.
\end{align}

When $r$ is large (that is, large relative to the Schwarzschild-radius $r_S=2G_0M$ for a source mass $M$), the metric coefficients become
\begin{align}
A&\simeq B^{-1},\label{eq:A}\\
B&\simeq 1-\frac{2G_0M}{r}.\label{eq:B}
\end{align}

This is a standard result of MOG which is obtained from the MOG field equations, solved in the presence of appropriately chosen, physically motivated initial conditions \cite{Moffat2007e}.

\subsection{Motion in a Spherically Symmetric Field}

To develop an equation of motion for a point particle and use it to derive a formula for light bending, we follow the approach presented by~\cite{Weinberg1972}. In particular, treating the photon as an ultrarelativistic point particle allows us to avoid postulating an action for the electromagnetic field that works in conjunction with the MOG gravitational action; this topic is the subject of on-going research.

We begin with the point particle action in MOG, which is written in the form
\begin{equation}
\begin{array}{lll}
S_\mathrm{TP}=&-\int(m+\alpha
q_5\phi_\mu u^\mu)~d\tau\\ [1em]
=&-\int(m\sqrt{g_{\alpha\beta}u^\alpha u^\beta}+\alpha
q_5\phi_\mu u^\mu)~d\tau,\label{eq:TPL}
\end{array}
\end{equation}
where $m$ is the point particle mass, $q_5$ is its fifth force charge, and $u^\alpha=dx^\alpha/d\tau$ is its four-velocity. The~dimensionless coefficient $\alpha$ is given by
\begin{equation}
\alpha=\frac{G_0-G_N}{G_N}=\left(\frac{G_\infty}{G_N}-1\right)\frac{M}{(\sqrt{M}+E)^2}.\label{eq:alpha}
\end{equation}

The fifth force charge is assumed to be proportional to $m$, such that $q_5=\kappa m$. Using the variational principle, we obtain the equations of motion in the form
\begin{equation}
m\left(\frac{du^\nu}{d\tau}+\Gamma_{\alpha\beta}^\nu u^\alpha u^\beta\right)=\alpha q_5
u^\beta g^{\nu\alpha}B_{\alpha\beta}.
\end{equation}

Integration yields
\begin{equation}
r^2\frac{d\phi}{d\tau}=J,\label{eq:dphidtau}
\end{equation}
and
\begin{equation}
A\left(\frac{dr}{d\tau}\right)^2+\frac{J^2}{r^2}-\frac{(1-\alpha\kappa
\phi_t)^2}{B}=-{\cal E},\label{eq:E}
\end{equation}
where ${\cal E}$ is another constant of integration.

From (\ref{eq:E}), taking the large-$r$ limit, we get
\begin{equation}
{\cal E}=1-v^2,
\end{equation}
where $v=dr/d\tau\simeq dr/dt$ is the velocity of the particle at infinity. For ultrarelativistic particles, ${\cal E}\rightarrow 0$.

For a non-relativistic particle, $v^2\ll 1$. Further, in the weak field limit, $1-2GM/r\simeq 1$, which leads to
\begin{equation}
\frac{d^2r}{dt^2}-\frac{J^2}{r^3}=\alpha\kappa
\phi_t'-\frac{GM}{r^2}.
\label{eq:nonrel}
\end{equation}

This equation of motion is a direct consequence of the solution of the field equations presented in Section~\ref{sec:formulation}, used in combination with the postulated test particle action (\ref{eq:TPL}).

The nonrelativistic equation of motion (\ref{eq:nonrel}) can be further simplified when only radial motion is considered, such that $J=0$:
\begin{equation}
\ddot{r}=-\left[1+\alpha-\alpha(1+\mu r)e^{-\mu r}\right]\frac{G_NM}{r^2}.\label{eq:ddot}
\end{equation}

This equation formed the basis of much of our investigation to date, including our successful modeling of the detailed rotation curves of spiral galaxies \cite{Brownstein2006a,Brownstein:2009gz}.

\subsection{Flat Rotation Curves and Velocity Dispersion}

MOG yields a regime of rotation curves significantly flatter than the Keplerian values in the spherically symmetric field of a point source at distances of $r\sim\mu^{-1}$. This effect arises as the weakening of the gravitational field with increasing radial distance is partially canceled out by the exponential vanishing of the repulsive Yukawa component. The rotational velocity $v$ for a circular orbit can be calculated using basic kinematics: $v^2/r = -\ddot{r}$, where $\ddot{r}$ is the radial acceleration given by Equation~(\ref{eq:ddot}). Thus we have
\begin{equation}
v=\sqrt{[1+\alpha-\alpha(1+\mu r)e^{-\mu r}]\frac{G_NM}{r}}.
\end{equation}

At $r=\mu^{-1}$, we obtain
\begin{equation}
v=\sqrt{[1+\alpha(1-2e^{-1})]G_NM\mu}.
\end{equation}

From Equation~(\ref{eq:mu}), we have $\mu=D/\sqrt{M}$ and we get
\begin{equation}
v=\sqrt{[1+\alpha(1-2e^{-1})]DG_N}\sqrt[4]{M},\label{eq:vrot}
\end{equation}
in agreement with the observationally established Tully-Fisher relation \cite{TF1977}, $v^4\appropto M$.

We expect the line-of-sight velocity dispersion $\sigma_{||}$ of an observed galaxy to be proportional to its typical rotational velocity. This justifies the use of the empirical Faber-Jackson relation \cite{FJ1976} for elliptical galaxies, which can be written as \cite{BT2008}:
\begin{equation}
\sigma_{||}\simeq 150\left(\frac{L}{10^{10} L_\odot}\right)^{1/4}~{\rm km}/{\rm s},\label{eq:FJ}
\end{equation}
where $L\propto M$ is the luminosity of the galaxy.

If Equation (\ref{eq:FJ}) represents the typical velocity dispersion of a spherically symmetric elliptical galaxy, it can also be viewed as the typical velocity dispersion of a randomly oriented disk galaxy. For a disk galaxy that is viewed edge-on, the velocity dispersion will be $\sqrt{3/2}$ times this value. This also represents a limiting case: the observed line-of-sight velocity dispersion cannot be larger for a virialized galaxy that is not disturbed by external (e.g., tidal) gravitational forces or non-gravitational interactions.

\subsection{The Bending of Light}

To calculate light bending, we return to the exact result in Equation~(\ref{eq:E}). We eliminate $d\tau$ using Equation~(\ref{eq:dphidtau}):
\begin{equation}
\frac{A}{r^4}\left(\frac{dr}{d\phi}\right)^2+\frac{1}{r^2}=-\frac{{\cal E}}{J^2}+\frac{(1-\alpha\kappa
\phi_t)^2}{J^2B}.\label{eq:drdphi}
\end{equation}

For an extreme astronomical lensing scenario, we consider lensing by a $10^{12}~M_\odot$ object with an impact parameter of only 10~kpc. In this case, $\alpha\kappa
\phi_t\simeq -8\times 10^{-5}$, which is negligible. (For comparison, for starlight grazing the Sun, $\alpha\kappa
\phi_t\simeq -6.4\times 10^{-14}$.) This justifies taking, in the weak field limit, $1-\alpha\kappa
\phi_t\simeq 1$ for an ultrarelativistic particle.

Solving for $\Delta\phi_\gamma=2|\phi-\phi_\infty|-\pi$ for photons in the weak field limit, we get
\begin{equation}
\Delta\phi_\gamma=2\left|\int_{r_0}^\infty{\frac{1}{r}\left[\frac{r^2}{r_0^2}\frac{B(r_0)}{AB}-\frac{1}{A}\right]^{-1/2}}~dr\right|-\pi.
\end{equation}

\textls[-15]{This formula is formally identical to the light bending formula in the weak field limit of general relativity, with one notable difference: instead of $G=G_N$, we are using $G=G_0=(1+\alpha)G_N$ in the Schwarzschild coefficients $A$ and $B$. From this formula, the approximate deflection can be calculated as \cite{Moffat2006a,Moffat2008a,Weinberg1972}:}
\begin{equation}
\Delta\phi_\gamma=\frac{4GM}{r_0}=\frac{4(1+\alpha)G_NM}{r_0}.\label{eq:bend}
\end{equation}

Thus, the bending of light under MOG is equal to the bending of light under general relativity using an effective (lensing) mass
\begin{equation}
M_L=(1+\alpha)M.\label{eq:effM}
\end{equation}

As $\alpha$ is given as a function of $M$, this equation can be solved for $M$. Specifically:
\begin{equation}
M_L=\left(1+\frac{G_\infty-G_N}{G_N}\frac{M}{(\sqrt{M}+E)^2}\right)M.\label{eq:GMMOG}
\end{equation}

When no dark matter is present, $M=M_b$, where $M_b$ is the baryonic mass. Using the value of $M_L$ determined from lensing observations, Equation~(\ref{eq:GMMOG}) can be used to solve for $M_b$.

\section{Einstein Rings}
\label{sec:rings}

When light from a distant source is deflected by a massive object, an observer aligned with the distant source and the massive object may see a ring-like image of the distant source. Such images are often referred to as Einstein rings.

If the distances to the remote source and to the massive object are known, the apparent size of the Einstein ring provides a direct means to estimate the effective gravitational mass of the massive object. In~the case of the dark matter model, this effective gravitational mass is the sum of the masses of baryonic matter and exotic dark matter. In the case of a modified gravity such as MOG, in the absence of exotic dark matter, the effective gravitational (lensing) mass is the mass of baryonic matter, enhanced by modified gravity effects.

Specifically, the apparent radius $\theta_E$ of an Einstein ring is given by, in radians \cite{Weinberg2008}:
\begin{equation}
\theta_E=\sqrt{\frac{4G_NM_L}{c^2}\frac{d_{LS}}{d_Ld_S}},\label{eq:thetaE}
\end{equation}
where $d_L$, $d_S$, and $d_{LS}$ are the angular diameter distances (related to the comoving distance $\xi$ by \mbox{$d=\xi/(1+z)$} in the case of a flat, $\Omega=1$ cosmology) to the lensing object, the source, and between the lensing object and the source, respectively. For the dark matter theory, the lensing mass is the sum of baryonic and dark matter masses: $M_L=M=M_b+M_{\rm DM}$. For MOG, the lensing mass is given by Equation~(\ref{eq:effM}), with~$M=M_b$. Given $\theta_E$, $d_S$, $d_L$, and $d_{LS}$, Equations~(\ref{eq:GMMOG}) and (\ref{eq:thetaE}) allow us to solve for $M_b$.

Together, Equations~(\ref{eq:FJ}) and (\ref{eq:GMMOG}) provide a testable relationship between the velocity dispersion $\sigma_{||}$ and the lensing mass $M_L$. The lensing mass can be estimated using Equation~(\ref{eq:thetaE}) from Einstein ring observations, whereas the velocity dispersion of the lensing object can be observed directly.

\subsection{Some Strong Lensing Candidates}

\begin{figure}[H]
\centering\includegraphics[width=0.65\linewidth]{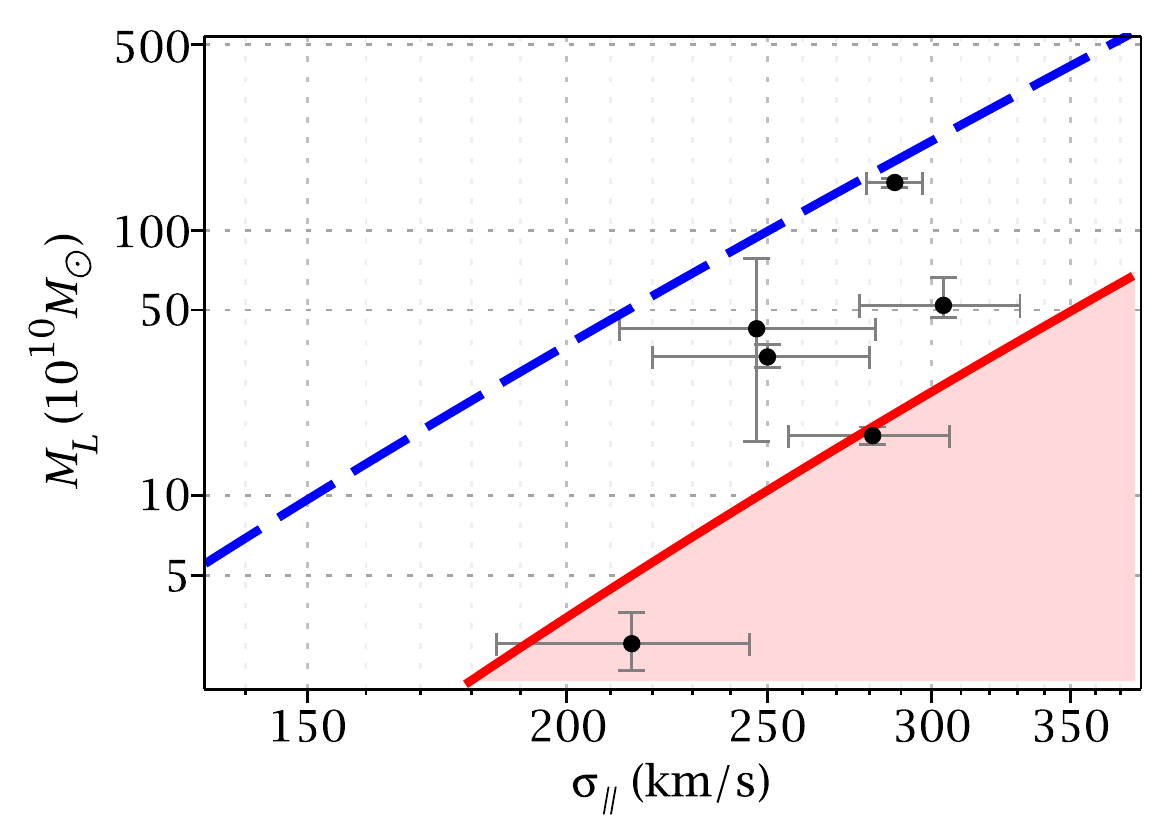}
\caption{The relationship between lensing masses and velocity dispersions. The six data points represent the six galaxies from Table~\ref{tb:lenses} for which velocity dispersion data are available. For these galaxies, the average mass-to-light ratio, computed from the lensing mass using MOG
	, is $\Upsilon_{\rm avg}=0.77$. The blue dashed line shows the Faber-Jackson relationship for this value. The lowest mass-to-light ratio in this sample of six galaxies is $\Upsilon_{\rm min}=0.24$. This case, calculated for an edge-on disk galaxy and indicated by the solid red line, represents a limiting case. No galaxy that is virialized and unaffected by external fields should be located in this region.}
\label{fig:masses6}
\end{figure}

For a small set of lensing galaxies, observational data for both $M_L$ and $\sigma_{||}$ are available. \cite{Leier2011} studied 21 strong lensing candidates, for which they estimated $M_L$ from the size of the lensed images. The velocity dispersion of six of these galaxies has been measured \cite{Warren1998,Tonry1999,Tonry1998,Koopmans2003,Koopmans2002,Foltz1992}. The lensing masses and velocity dispersions for these six objects are included in Table~\ref{tb:lenses}, and plotted in Figure~\ref{fig:masses6}.

It appears that these six lensing objects do not follow any clearly discernible trend. This is perhaps to be expected, given the difficulty of measuring the velocity dispersions of such very distant objects. On the other hand, none of the six lenses have velocity dispersions that would place them more than $1\sigma$ in the ``forbidden region'' (shaded area in Figure~\ref{fig:masses6}), where the velocity dispersion would be excessive compared to the lensing mass.

\begin{table}[H]
\caption{Data for 21 strong lensing objects. Lensing mass and stellar luminosity estimates are from \cite{Leier2011}. The value of the baryonic mass-to-light ratio under MOG, $\Upsilon_{\rm MOG}$, is predicted, as described in the text.\label{tb:lenses}}
\centering
\begin{tabular}{lcccccc}\toprule
\multirow{2}{*}{{\bf Lens}}&\boldmath{$M_L$}&\boldmath{$L_V$}&\boldmath{$\sigma_{||}$}&\multirow{2}{*}{\boldmath{$\Upsilon_{\rm MOG}$}}\\
&\boldmath{$(10^{10}~M_\odot)$}&\boldmath{$(10^{10} L_\odot)$}&{\bf (km/s)}&\\\midrule
Q0047   &$\phantom{0}33.27^{+3.80\phantom{0}}_{-3.03\phantom{0}}$&$2.30^{+0.59}_{-0.47}$&250(30) $^1$\phantom{}&$0.97_{-0.20}^{+0.25}$\\[.5em]
Q0142   &$\phantom{0}45.81^{+3.31\phantom{0}}_{-5.50\phantom{0}}$&$4.59^{+2.32}_{-1.61}$&---                  &$0.65_{-0.22}^{+0.35}$\\[.5em]
MG0414  &$\phantom{}102.76^{+16.55\phantom{}}_{-19.04\phantom{}}$&$3.54^{+2.01}_{-1.31}$&---                  &$1.74_{-0.63}^{+1.02}$\\[.5em]
B0712   &$\phantom{0}16.13^{+2.55\phantom{0}}_{-2.04\phantom{0}}$&$1.67^{+0.50}_{-0.40}$&---                  &$0.71_{-0.17}^{+0.22}$\\[.5em]
HS0818  &$\phantom{0}36.88^{+3.69\phantom{0}}_{-5.48\phantom{0}}$&$1.74^{+0.57}_{-0.44}$&---                  &$1.40_{-0.34}^{+0.48}$\\[.5em]
RXJ0911 &$\phantom{0}73.04^{+1.15\phantom{0}}_{-1.36\phantom{0}}$&$4.85^{+2.47}_{-1.69}$&---                  &$0.93_{-0.31}^{+0.50}$\\[.5em]
BRI0952 &$\phantom{0}14.76^{+5.67\phantom{0}}_{-5.87\phantom{0}}$&$1.79^{+0.57}_{-0.44}$&---                  &$0.61_{-0.15}^{+0.20}$\\[.5em]
Q0957   &$\phantom{}151.29^{+5.67\phantom{0}}_{-6.58\phantom{0}}$&$8.29^{+2.51}_{-1.95}$&288(9) $^2$\phantom{0}&$1.06_{-0.25}^{+0.33}$\\[.5em]
LBQS1009&$\phantom{0}64.73^{+11.97\phantom{}}_{-24.46\phantom{}}$&$3.09^{+1.86}_{-1.20}$&---                  &$1.31_{-0.50}^{+0.83}$\\[.5em]
B1030   &$\phantom{0}55.09^{+2.11\phantom{0}}_{-3.52\phantom{0}}$&$4.61^{+2.24}_{-1.58}$&---                  &$0.76_{-0.25}^{+0.39}$\\[.5em]
HE1104  &$\phantom{0}72.80^{+3.68\phantom{0}}_{-3.22\phantom{0}}$&$4.11^{+1.82}_{-1.29}$&---                  &$1.09_{-0.33}^{+0.50}$\\[.5em]
PG1115  &$\phantom{0}16.80^{+1.31\phantom{0}}_{-1.23\phantom{0}}$&$1.52^{+0.94}_{-0.62}$&281(25) $^3$\phantom{}&$0.81_{-0.31}^{+0.55}$\\[.5em]
B1152   &$\phantom{0}30.43^{+4.23\phantom{0}}_{-5.86\phantom{0}}$&$3.02^{+0.92}_{-0.72}$&---                  &$0.68_{-0.16}^{+0.22}$\\[.5em]
B1422   &$\phantom{00}5.72^{+0.39\phantom{0}}_{-0.24\phantom{0}}$&$0.46^{+0.17}_{-0.13}$&---                  &$1.11_{-0.31}^{+0.44}$\\[.5em]
SBS1520 &$\phantom{0}41.98^{+1.49\phantom{0}}_{-1.82\phantom{0}}$&$4.37^{+2.64}_{-1.69}$&---                  &$0.63_{-0.24}^{+0.40}$\\[.5em]
B1600   &$\phantom{0}16.38^{+0.82\phantom{0}}_{-1.67\phantom{0}}$&$0.62^{+0.22}_{-0.16}$&---                  &$1.94_{-0.50}^{+0.69}$\\[.5em]
B1608   &$\phantom{0}42.49^{+35.50\phantom{}}_{-26.52\phantom{}}$&$11.5^{+3.40}_{-2.67}$&247(35) $^4$\phantom{}&$0.24_{-0.06}^{+0.07}$\\[.5em]
MG2016  &$\phantom{0}52.05^{+14.23\phantom{}}_{-5.13\phantom{0}}$&$7.12^{+6.58}_{-3.75}$&304(27) $^5$\phantom{}&$0.47_{-0.22}^{+0.52}$\\[.5em]
B2045   &$\phantom{}173.07^{+21.02\phantom{}}_{-34.20\phantom{}}$&$5.76^{+2.01}_{-1.50}$&---                  &$1.74_{-0.45}^{+0.60}$\\[.5em]
HE2149  &$\phantom{0}26.82^{+2.07\phantom{0}}_{-2.19\phantom{0}}$&$5.42^{+2.12}_{-1.57}$&---                  &$0.34_{-0.10}^{+0.14}$\\[.5em]
Q2237   &$\phantom{00}2.76^{+0.85\phantom{0}}_{-0.58\phantom{0}}$&$0.26^{+0.14}_{-0.10}$&215(30) $^6$\phantom{}&$1.08_{-0.39}^{+0.61}$
\\\bottomrule
\end{tabular}\\
\begin{tabular}{lcccccc}
\multicolumn{7}{l}{\footnotesize
$^1$\cite{Warren1998};
$^2$\cite{Tonry1999};
$^3$\cite{Tonry1998};
$^4$\cite{Koopmans2003};
$^5$\cite{Koopmans2002};
$^6$\cite{Foltz1992}.}
\end{tabular}
\end{table}

\subsection{Predicting Mass-to-Light Ratios}

The luminosity of a distant galaxy can be obtained by optical observation. On the other hand, the~mass of a distant galaxy usually cannot be measured: it must be inferred from the galaxy's observable properties using an applicable model of galaxy evolution. The predicted amount of baryonic matter in a galaxy, therefore, depends strongly on the choice of model and may vary significantly between models.

On the other hand, when a galaxy is a strong lensing object, its effective mass can be calculated as per our preceding discussion, by observing the size of the Einstein ring and using the redshifts of the lensing and lens objects as proxies for distance.

Under the MOG theory, the effective lensing mass is due entirely to the baryonic content of the galaxy and the variable gravitational constant that is associated with it; as such, the baryonic mass $M_b$ can be estimated using Equation~(\ref{eq:GMMOG}). Given the known luminosities $L$ of the lensing objects, the ratio of $\Upsilon_{\rm MOG}=M_b/L$ can be then calculated. We present the predicted baryonic mass-to-light ratios for all 21 lensing objects listed in Table~\ref{tb:lenses}. The average for the 21 listed galaxies if $\Upsilon_{\rm avg}=0.96$. In comparison, the~baryonic mass-to-light ratio in the disk of the Milky Way is $\Upsilon_V = 1.5\pm 0.2$ \cite{Flynn2006}. Our results from MOG imply that at higher redshifts galaxies may have lower  mass-to-light ratios. This is in agreement with the evolution of the mass-to-light ratio in galaxies. At higher redshifts the mass function is dominated by massive stars and the result is a lower $\Upsilon$; this is confirmed by observation \cite{Rusin2003}.

\section{Abell~520 and the Bullet Cluster}
\label{sec:clusters}

Large clusters of galaxies and, in particular, collisions of such clusters have been observed in recent years and formed the subject of much discussion concerning the validity of the standard $\Lambda$CDM model of~cosmology.

In the CDM scenario, when two large clusters collide, much of the gas present in the galaxies slows down and heats up. This was indeed observed using X-ray telescopy, which detected X-ray radiation from large quantities of hot ($\sim$$10^7$~K) gas at the locus of the collision. Stars and CDM halos of galaxies, on the other hand, pass through each other in a collisionless manner and continue along the original trajectories of the colliding clusters, their path modified only by gravitational interactions.

When background objects are viewed through colliding clusters, significant weak lensing should be observable. The lensing is proportional to the gas mass at the locus of the collision. Lensing proportional to the combined stellar and CDM mass (a sum that is dominated by the CDM mass) will be seen at the locus of the visible galaxies.

In particular, the case of the Bullet Cluster 1E0657-558 has been viewed by many as a strong indication in favor of the CDM scenario, since the visual brightness of the post-collision galaxies, combined with any sensible mass-to-light ratios, do not even come close to yielding the mass that would be necessary to produce the observed lensing. In contrast, the colliding clusters Abell~520 indicate weak lensing that, under the standard CDM scenario, can only be explained if significant quantities of CDM were present in the object's core. This is at odds with the standard assumption that the CDM medium is truly collisionless, interacting only gravitationally. Cold dark matter, like stars, would have continued without collisions and would be located where galaxies are currently visible. In actuality, no visible galaxies are observed near the core of Abell~520 \cite{Jee2012}.

\subsection{Applying MOG to Colliding Clusters}

As we have seen in the previous sections, much of the phenomenology of the MOG theory is captured by the concept of an effective lensing mass, as given by Equation~(\ref{eq:effM}). The lensing mass $M_{\rm L}$ determines both the bending of light and low-velocity, weak field dynamics far from a source (nearer the source, the repulsive vector force of the theory must also be taken into account). The value of $\alpha$, given by Equation~(\ref{eq:alpha}), ranges from near zero for small objects (e.g., a single star) all the way up to $\sim$$19$ for very large, compact gravitational sources.

This phenomenology has been developed for single, isolated, point-like sources of gravity, and~cannot be applied trivially to extended distributions of matter, as the MOG theory is inherently nonlinear. Nonetheless, we expect this phenomenology to remain valid, at least approximately and as an upper limit for $\alpha$, for extended objects such as large clusters of galaxies.

Assuming the validity of the MOG theory, the value of $\alpha$ can be estimated from observation. Consider an object with known dynamical (lensing) mass $M_L$ and luminosity $L$. Assuming a stellar mass-to-light ratio $\Upsilon_\star$ (which may be determined using standard galaxy evolution models) yields an estimated stellar mass $M_\star=\Upsilon_\star L$. The amount of gas present is characterized by the gas fraction, $f_g$, which relates the lensing mass to the mass of gas $M_g$ deduced from direct observation, and is defined as
\begin{equation}
f_g=\frac{M_g}{M_L}.
\end{equation}

Together, the gas and stellar mass yield the baryonic (non-exotic) mass:
\begin{equation}
M_b=M_g+M_\star = f_gM_L+\Upsilon_\star L.
\end{equation}

In the MOG theory, the effective mass is $(1+\alpha)$ times the baryonic mass (stars and gas) with no exotic dark matter present. This allows us to determine $\alpha$ as
\begin{equation}
\alpha=\frac{M_L}{M_b}-1.
\end{equation}

This determination of $\alpha$ can be contrasted with the values of $\alpha$ that we can calculate directly from either the lensing mass $M_L$ or the baryonic mass $M_b$ using Equations~(\ref{eq:effM}) and (\ref{eq:GMMOG}).

The Bullet Cluster 1E0657-558 has been studied in detail from the MOG perspective: it has been shown \cite{Brownstein2007} that the weak lensing map associated with this cluster is consistent with the MOG prediction in the absence of exotic dark matter. In the following, we review the case of the Bullet Cluster and also apply our reasoning to the case of Abell~520.

\subsection{The Bullet Cluster 1E0657-558}

The Bullet Cluster has been studied extensively by \cite{Markevitch2006} and others. In particular, they determined the dynamical mass of the two visible cluster components (``main cluster'' and ``subcluster'') from weak lensing and compared these to the visual magnitudes of these components.

In the $B$-band, the mass-to-light ratio for the main cluster was found to be in the range 275--314 for the main cluster and 271--297 for the subcluster, depending on the choice of observational methodology~\cite{Clowe2004}. In combination with the lensing masses and gas fractions estimated by~\cite{Markevitch2006}, these values allow us to estimate the MOG $\alpha$ parameter, as shown in Table~\ref{tb:BC}. For the purposes of this estimate, we used a stellar mass-to-light ratio of $\Upsilon_\star=1$; however, the final value is not particularly sensitive to $\Upsilon_\star$ due to the relative smallness of the stellar component in the overall mass.

We can compare these estimates of $\alpha$ with the values of $\alpha$ obtained from the lensing masses. For the main cluster, subcluster, and main gas peak, we obtain the values of $\alpha(M_L)$ that are given in Table~\ref{tb:BC}.

\begin{table}[H]
\caption{{Lensing mass}, gas fraction, stellar luminosity and the MOG $\alpha$ parameter (estimated assuming a stellar mass-to-light ratio of $\Upsilon_\star=1$) and the same parameter computed from $M_L$ for components of the Bullet Cluster. Using data from \cite{Clowe2004} and \cite{Markevitch2006} (for the main gas peak).\label{tb:BC}}
\begin{centering}
\begin{tabular}{lccccc}\toprule
\multirow{2}{*}{{\bf Component}}&\boldmath {$M_L$}&\multirow{2}{*}{\boldmath {$f_g$}}&\boldmath {$L_\star$}&\multirow{2}{*}{\boldmath {$\alpha$}}&\multirow{2}{*}{\boldmath {$\alpha(M_L)$}}\\
&\boldmath {${10^{13}M_\odot}$}&&\boldmath {$10^{11}L_\odot$}&&\\\midrule
Main cluster&9.5(1.5)&0.09(1)&3.5&$\phantom{0}9.7^{+1.3}_{-1.1}$&17.7\\[.5em]
Subcluster  &6.6(1.9)&0.04(1)&2.1&$\phantom{}22.2^{+7.6}_{-4.8}$&17.5\\[.5em]
Gas peak&10.8(6)&0.19(3)&N/A&$\phantom{0}4.3^{+1.0}_{-0.8}$&18.6\\\bottomrule
\end{tabular}\par
\end{centering}
\end{table}

These idealized values for $\alpha$ characterize point sources. As expected, the values of $\alpha$ obtained by comparing observational estimates of the lensing mass versus baryonic mass are smaller. We attribute this to the fact that the components of the Bullet Cluster are themselves extended objects spread over several hundred kpc. By comparing the ideal values of $\alpha$ to the values obtained from observation, we find that they come closest for the subcluster. This is consistent with the visual observation that the subcluster is smaller, more compact than its main counterpart. Similarly, the fact that the value of $\alpha$ obtained from observation is smallest for the gas peak is indicative of the fact that the gas peak corresponds to a relatively diffuse cloud of gas, not compact galaxies of stars.

\subsection{Applying MOG to Abell~520}

\begin{figure}[H]
  \centering
  \includegraphics[width=0.65\linewidth,angle=-90]{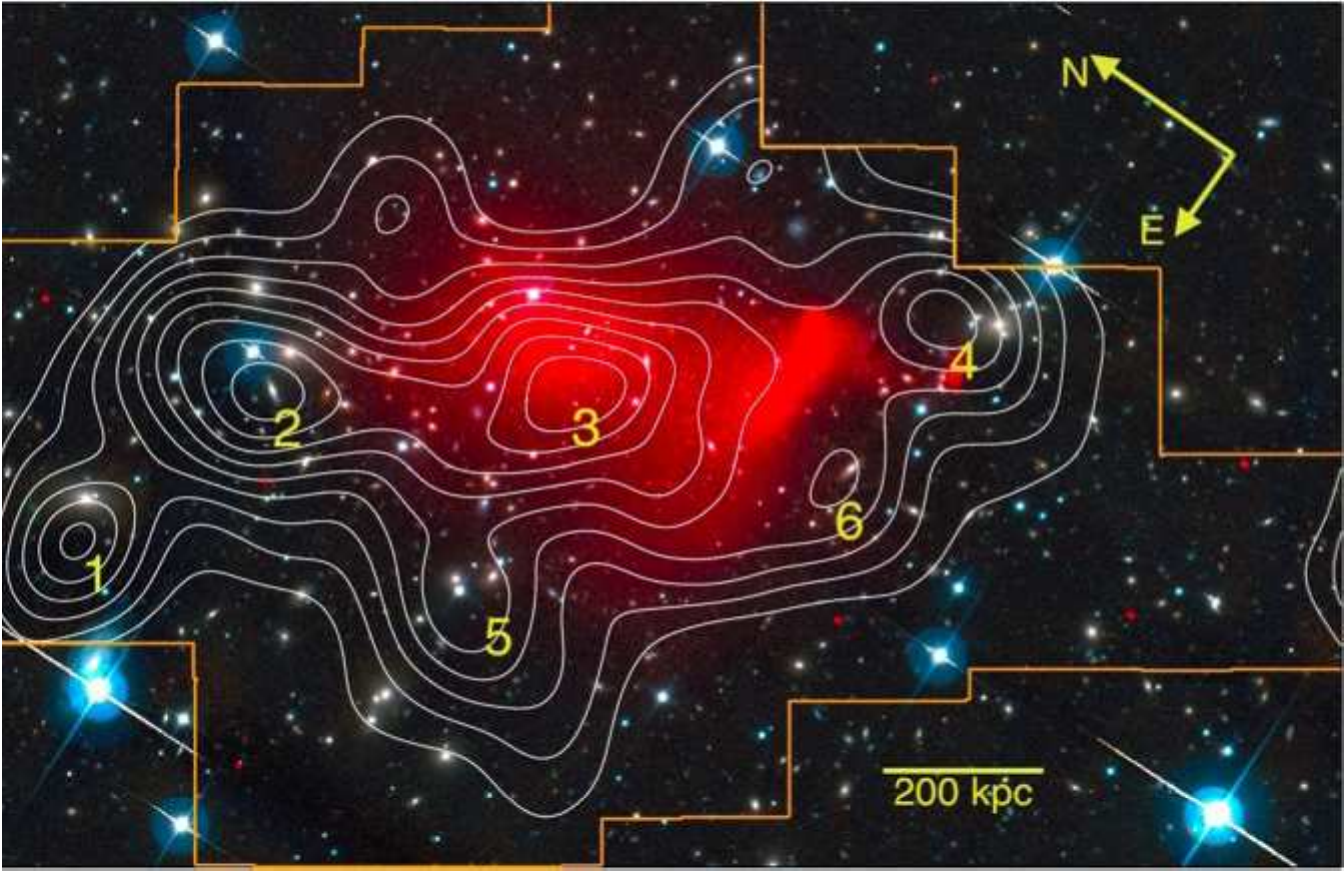}
\caption{The Abell~520 cluster with distribution of galaxies, gas and contours from weak lensing \cite{Jee2012}.}
  \label{fig:A520}
\end{figure}

We now turn our attention to the colliding clusters Abell~520. Weak lensing by the Abell~520 cluster was recently explored using data from the Wide Field Planetary Camera of the Hubble Space Telescope~\cite{Jee2012}. The authors conclude that this cluster has a dark core that coincides with the peak X-ray luminosity (see Figure~\ref{fig:A520}) and that there are no visible galaxies in this central~core.

This cluster has a main structure labeled P3 and substructures labeled P1, P2, P4, P5 and P6. Visible matter in the central main substructure is mainly gas while the other substructures have a mixture of gas and galaxies.

Using the parameters of the lensing cluster as the projected mass, luminosity of galaxies in B-band, mass to light ratio and the fraction of gas in each subcluster, we can find the MOG $\alpha$ parameter for each of the substructures (see Table~\ref{tb:A520}).

As in the case of the Bullet Cluster, we find that the values of $\alpha$ estimated this way are below the values estimated from the lensing mass alone, treating the substructures as point sources. This is just as expected: for extended objects like these substructures, the point source estimate should serve as an upper~limit.

Furthermore, just as we would expect, the values of $\alpha$ computed from observation are lower for substructures that have a higher gas content. This coincides with our interpretation that gas structures are diffuse and extended, whereas galaxies are more compact.

\begin{table}
\caption{Lensing mass, gas fraction, stellar luminosity and the MOG $\alpha$ parameter (estimated assuming a~stellar mass-to-light ratio of 1) and the same parameter computed from $M_L$ for components of Abell~520~\cite{Jee2012}. Note that P3 corresponds to the ``dark core''.\label{tb:A520}}
\begin{centering}
\begin{tabular}{lccccc}\toprule
\multirow{2}{*}{{\bf Component}}&\boldmath {$M_L$}&\multirow{2}{*}{\boldmath {$f_g$}}&\boldmath {$L_B$}&\multirow{2}{*}{\boldmath {$\alpha$}}&\multirow{2}{*}{\boldmath {$\alpha(M_L)$}}\\
&\boldmath {($10^{13}M_\odot$)}&&\boldmath {($10^{11}L_\odot$)}&&\\\midrule
P1            &2.63(48)&$<$0.06&1.54&$>$13.8&18.2\\
P2            &3.83(42)&$<$0.08&3.58&$>$9.7&18.3\\
P3            &4.00(38)&$<$0.14&0.68&$>$6.0&18.4\\
P4            &3.64(45)&$<$0.08&2.95&$>$9.9&18.3\\
P5            &3.02(40)&$<$0.05&2.12&$>$15.5&18.3\\
P6            &3.33(40)&$<$0.06&1.23&$>$14.3&18.3\\\bottomrule
\end{tabular}\par
\end{centering}
\end{table}

\section{Conclusions and Outlook}
\label{sec:conclusions}

In this paper, we studied the predictions of the MOG modified theory of gravity as applied to weak and strong gravitational lensing.

For strong lensing objects, MOG can be used to predict a relationship between velocity dispersions and lensing masses. This is indeed the same derivation that allows us to use MOG to derive the Tully-Fisher relationship. We found that given the small number of lensing objects for which reliable velocity dispersion data are available in the literature, all are consistent with the predictions and constraints of the MOG~theory.

We also applied MOG to the case of the merging clusters 1E0657-558 (the Bullet Cluster) and Abell~520. For the former, a detailed study was conducted in the past \cite{Brownstein2007}, showing excellent agreement between weak lensing observations and the MOG prediction. Here, we demonstrated why this is so: that the effective lensing masses of specific regions of the Bullet Cluster are consistent with their luminosities and mass-to-light ratios under the MOG theory, in the absence of dark matter. We then applied the same reasoning to the merging clusters Abell~520. Under the standard cold dark matter scenario, weak lensing of this cluster can only be explained by assuming the presence of a dark matter core, which is inconsistent with the collisionless nature of the cold dark matter component and the lack of visible galaxies in this core region. For MOG, no such difficulty exists: as in the case of the Bullet Cluster, the lensing masses and luminosities are consistent with the predictions of the theory.

In these studies, we applied the MOG point source solution to the objects that were examined. This~necessarily introduced a degree of uncertainty in the predictions: due to the inherently nonlinear nature of the MOG theory, the point source solution can only be considered as a limiting case when spatially extended, diffuse objects are examined. An alternative is to use a density profile and the generalized MOG Poisson equation, as it was done in \cite{Brownstein2007,Israel2016}. Apart from a difference in the predicted value of $\alpha$ (the~point-source solution in the current paper yielded a larger value, as expected) both approaches yield consistent results.

\acknowledgments{ J. W. Moffat thanks the John Templeton Foundation for generous financial support. The research was partially supported by National Research Council of Canada. Research at the Perimeter Institute for Theoretical Physics is supported by the Government of Canada through NSERC and by the Province of Ontario through the Ministry of Research and Innovation (MRI).}

\authorcontributions{The authors contributed equally to this work.}%

\conflictsofinterest{The authors declare no conflict of interest.}

\reftitle{References}

\end{document}